\documentclass[11pt]{article}             % Use with LaTeX2e
\usepackage{osid}                         %
\usepackage{amsmath}
\usepackage{amsfonts}
\usepackage{amssymb}
\usepackage{graphics}
\usepackage{graphicx}

\title{Effective Hamiltonians for Complexes of Unstable Particles}
\author{Krzysztof Urbanowski \\{\footnotesize\it University of Zielona G\'{o}ra, Institute of Physics, %\\
ul. Prof. Z. Szafrana 4a,\\ 65--516 Zielona G\'{o}ra, Poland. \& E-mail: K.Urbanowski@proton.if.uz.zgora.pl}\\[2ex]
}

\begin{document}

\maketitle
\begin{abstract}
Effective Hamiltonians   governing the time evolution
in a subspace of unstable states can be found using
more or less accurate approximations. A convenient tool
for deriving them is the evolution equation for a
subspace of state space sometime called the
Krolikowski-Rzewuski (KR) equation. KR equation results
from the Schr\"{o}dinger equation for the total system
under considerations. We will discuss properties of approximate effective Hamiltonians
derived using KR equation for $n$--particle, two particle and for one particle subspaces.
In a general case these affective Hamiltonians depend on time $t$.
We show that at times much longer than times at which the exponential decay take place
the real part of the exact effective Hamiltonian for the one particle subsystem (that is
the  instantaneous energy) tends to the minimal energy of the total system when
$t \rightarrow \infty$ whereas the imaginary part of this effective Hamiltonian
tends to the zero as $t\rightarrow \infty$.
\end{abstract}

\noindent
\section{Introduction}

The standard
approach to searching for the properties of subsystems of unstable particles
makes use of more or less accurate approximate methods to solve evolution
equation for such subsystems. A typical example of such methods are
Weisskopf--Wigner (WW) approximation \cite{WW} or Lee--Oehme--Yang
(LOY) approximation \cite{Lee1,Comins}. All intermediate
steps of WW or LOY approximations leading to the final formulae describing
the time evolution of unstable particles are rather far from
mathematical precision. What is more, attempts to confront the
predicted properties of the considered systems, obtained within
the use of such approximate methods, with those following from the
analytical properties of the exact solutions of the quantum
evolution equation are rather sporadic. These analytical
properties can be extracted from properties of the transition amplitudes
\begin{equation}
A_{\alpha \beta}(t) = \langle \alpha |U(t)|\beta \rangle,
\label{A-cmst}
\end{equation}
where $|\alpha \rangle, |\beta\rangle \in {\cal H}$, $\cal H$ is
the Hilbert state space of the total system considered, and $U(t)$
is the total unitary evolution equation solving the Schr\"{o}dinger equation
\begin{equation}
i  \frac{\partial}{\partial t} U(t)|\psi \rangle = H U(t)|\psi
\rangle, \; \; U(0) = I, \label{Schrod}
\end{equation}
(we use $\hbar = c = 1$ units), $I$ is the unit operator in $\cal
H$, $|\psi \rangle \equiv |\psi ; t_{0} = 0\rangle$  $\in {\cal
H}$  is  the  initial  state of the system, (in  our case $|\psi
;t\rangle = U(t) |\psi \rangle$), and $H$ is the total
(selfadjoint) Hamiltonian, acting in $\cal H$.

Amplitudes $A_{\alpha \beta}(t)$ can be expressed in terms of the
energy (mass) densities $\omega_{\alpha \beta}(E)$ as follows
\begin{equation}
A_{\alpha \beta}(t) = \int_{{\rm Spec}(H)} \omega_{\alpha
\beta}(E)\, e^{\textstyle{-i Et}}\,dE, \label{cmst-rho}
\end{equation}
Assuming that the exact properties of real systems containing
unstable particles are described by the exact solutions of Eq.
(\ref{Schrod}), properties of amplitudes $A_{\alpha \beta}(t)$
following, eg., from symmetries of the system under considerations
can be used to examine properties of some parameters describing
the unstable particles and obtained by means of
$A_{\alpha \beta}(t)$ following from approximate methods of calculations.
A typical example is a subsystem of  neutral mesons and properties of
$A_{\alpha \beta}(t)$ following from $\cal CP$ or $\cal CPT$ invariance of $H$,
\cite{nowakowski}. Moreover amplitudes $A_{\alpha \beta}(t)$ are convenient for
numerical simulations of time evolution of the states considered:
It is sufficient to assume the form of the densities $\omega_{\alpha
\beta}(E)$ and then to use computer methods to find $A_{\alpha
\beta}(t)$ as a function of time $t$.

In the case of neutral kaons (neutral mesons in general)
all known properties including
CP--  and hypothetically  possible CPT--violation
effects in such complexes are  described by solving the
Schr\"{o}dinger--like evolution equation \cite{Lee1} --- \cite{improved},
\begin{equation}
i \frac{\partial}{\partial t} |\psi ; t \rangle_{\parallel} =
H_{\parallel} |\psi ; t \rangle_{\parallel} \label{l1}
\end{equation}
for $|\psi ; t \rangle_{\parallel}$ belonging to the subspace
${\cal H}_{\parallel} \subset {\cal H}$, e.g., spanned by
orthonormal neutral  kaons states $|K_{0}\rangle = |{\bf 1}\rangle, \;
|{\overline{K}}_{0}\rangle = |{\bf 2}\rangle\; \in {\cal H}$,
$\langle{\bf j}|{\bf k}\rangle = {\delta}_{jk}$, $j,k =1,2$,
(then states corresponding to the decay products belong to ${\cal H}
\ominus {\cal H}_{\parallel} \stackrel{\rm def}{=} {\cal
H}_{\perp}$), and nonhermitian effective Hamiltonian
$H_{\parallel}$ obtained usually by means of the
LOY approach (within the WW approximation) \cite{Lee1,Comins,improved},
\begin{equation}
H_{\parallel} \equiv M - \frac{i}{2} \Gamma, \label{new1}
\end{equation}
where $H_{\parallel}, M=M^{+}, \Gamma = \Gamma^{+}$ are $(2 \times 2)$ matrices,  $M$ is the mass matrix and $\Gamma$ is the decay matrix.

Within the WW approximation for a single particle subsystem the evolution equation
has a similar form to (\ref{l1}),
\begin{equation}
i \frac{\partial a(t)}{\partial t} = h_{WW}\; a (t), \label{WW-1}
\end{equation}
where $|\psi;t\rangle_{||} \equiv a(t)\,|\alpha\rangle$ and {\em dim}
${\cal H}_{||}=1$ , $a(0) =1$, and  $h_{WW} =$ \linebreak  $E^{0}_{\alpha}\, -
\, \frac{i}{2}\,\gamma^{0}_{\alpha}$ is the WW effective hamiltonian governing the time
evolution in one dimensional subspace of states ($E^{0}_{\alpha}$ is the energy of
the system  in the state $|\alpha\rangle$ and $\gamma^{0}_{\alpha}$ is the decay width). In this case the amplitude
$A_{\alpha \beta}(t)$, (\ref{cmst-rho}), can be replaced by $a(t) \stackrel{\rm def}{=}
A_{\alpha \alpha}(t)$ and $\omega_{\alpha \beta}(E)$ by $\omega(E)
\stackrel{\rm def}{=} \omega_{\alpha \alpha}(E)$, and here $\omega(E) \geq 0$ \cite{Fock}.

The analysis of the models of the decay processes shows that
the LOY and WW effective Hamiltonians appearing in (\ref{l1}) and (\ref{WW-1})
describe properties of two particle  or single particle complexes
to a very high accuracy  for a wide time range $t$: From $t$
suitably later than some $T_{0} \simeq t_{0}= 0$ but $T_{0} >
t_{0}$ up to $t \gg \tau = 1/{\gamma^{0}_{\alpha}}$
and smaller than the transition time $t = t_{as}$, where $t_{as}$ denotes the
time $t$ for which the nonexponential deviations of the survival probability
begin to dominate.

In \cite{Khalfin} assuming that the spectrum of $H$ must be bounded
from below, $(Spec.(H)\; = \;[E_{min}, +\infty)\; > \; -\infty)$, and using the Paley--Wiener
Theorem \cite{Paley} it was proved that in the case of unstable states there must be
\begin{equation}
|a (t)| \; \geq \; B\,e^{\textstyle - b \,t^{q}},
\label{|a(t)|-as}
\end{equation}
for $|t| \rightarrow \infty$. Here $B > 0,\,b> 0$ and $ 0 < q < 1$.
This means that the decay law ${\cal P}(t) = |a(t)|^{2}$ of unstable
states decaying in the vacuum can not be described by
an exponential function of time $t$ if time $t$ is suitably long, $t
\rightarrow \infty$, and that for these lengths of time ${\cal P}(t)$ tends to
zero as $t \rightarrow \infty$  more slowly than any exponential function of $t$ \cite{Fonda}.
From the model analysis it follows that in the general
case the decay law ${\cal P}(t)$ takes the inverse
power--like form $t^{- \lambda}$, (where $\lambda
> 0$), for suitably large $t \geq t_{as}\gg \tau$. Not long ago this effect
was confirmed experimentally: in the experiment described  in \cite{rothe},
the evidence of deviations from the exponential decay law at long times was
reported. The conclusion is that the LOY, WW and similar effective Hamiltonians
can not be used when one analysis a very long time properties of unstable systems.

The aim of this paper is to analyze more accurate approximations for
the effective Hamiltonians governing the time evolution in subspace of
unstable states  than those given LOY or WW  formulae and to analyze very long
time properties of the effective hamiltonian for an one particle subsystem.

\section{Beyond the WW and LOY approximations}
\subsection{Approximate formulae for $H_{\parallel}$ --- a general case}

The approximate formulae for $H_{\parallel} \equiv H_{\parallel}(t)$ have been  derived
in \cite{acta} --- \cite{10}  using  the  Krolikowski--Rzewuski equation   for
the projection of a state vector \cite{7}, which results from the
Schr\"{o}dinger  equation (\ref{Schrod}) for  the  total system
under consideration, and, in the  case  of the initial conditions
of the type $|\psi  \rangle  \equiv  |\psi  \rangle_{\parallel}, \;\;
|\psi\rangle_{\perp} = |\psi \rangle - |\psi\rangle_{\parallel} = 0  $,
 takes the following form
\begin{equation}
( i \frac{\partial}{ {\partial} t} - PHP ) U_{\parallel}(t)|\psi
\rangle_{||}
 =  - i \int_{0}^{\infty} K(t - \tau ) U_{\parallel}
( \tau )|\psi \rangle_{||} d \tau,   \label{KR1}
\end{equation}
where $ U_{\parallel} (0)  =  P$, $K(t)  =  {\mit \Theta} (t) PHQ\,
\exp \,[-itQHQ]\,QHP, \label{K} $ and ${\mit \Theta} (t)  =$
\linebreak $\{ 1 \;{\rm for} \; t \geq 0,  \;
0 \; {\rm for} \; t < 0 \}$ is the unit step function.

The integro--differential equation (\ref{KR1}) can be replaced by
the following differential one (see \cite{acta} --- \cite{7})
\begin{equation}
( i \frac{\partial}{ {\partial} t} - PHP - V_{||}(t) )
U_{\parallel}(t)|\psi \rangle_{||} = 0, \label{KR2}
\end{equation}
where
\begin{equation}
PHP + V_{||} (t) \stackrel{\rm def}{=} H_{||}(t). \label{H||=def}
\end{equation}
Taking into account (\ref{KR1}) and (\ref{KR2}) or (\ref{l1}) one
finds from (\ref{KR1})
\begin{equation}
V_{\parallel} (t) U_{\parallel} (t) = - i \int_{0}^{\infty} K(t -
\tau ) U_{\parallel} ( \tau ) d \tau \stackrel{\rm def}{=} - iK
\ast U_{\parallel} (t) . \label{V||=def}
\end{equation}
(Here the asterisk, $\ast$, denotes the convolution: $f  \ast g(t)
= \int_{0}^{\infty}\, f(t - \tau ) g( \tau  ) \, d \tau$
).\linebreak Next, using this relation and a retarded Green's
operator  $G(t)$ for the equation (\ref{KR1})
\begin{equation}
G(t) = - i {\mit \Theta} (t) \exp (-itPHP)P, \label{G}
\end{equation}
one obtains \cite{9,10}
\begin{equation}
U_{\parallel}(t) = \Big[ {\it 1}_{\ast} + \sum_{n = 1}^{\infty}
(-i)^{n}L \ast \ldots \ast L \Big] \ast U_{\parallel}^{(0)} (t) ,
\label{U||-szer}
\end{equation}
where $L$ is convoluted $n$ times, ${\it 1}_{\ast} \equiv {\it
1}_{\ast}(t) \equiv \delta (t)$, $L(t) = G \ast K(t) $, and
\begin{equation}
U_{\parallel}^{(0)} = \exp (-itPHP) \; P \label{U0}
\end{equation}
is a "free" solution of Eq. (\ref{KR1}). Thus from (\ref{V||=def})
\begin{equation}
V_{\parallel}(t) \; U_{\parallel}(t) = - i K \ast \Big[ {\it
1}_{\ast} + \sum_{n = 1}^{\infty} (-i)^{n}L \ast \ldots \ast L
\Big] \ast U_{\parallel}^{(0)} (t) , \label{V-szer}
\end{equation}
Of course, the  series (\ref{U||-szer}), (\ref{V-szer}) are convergent if
$\parallel L(t) \parallel < 1$. If for every $t \geq 0$
\begin{equation}
\parallel L(t) \parallel \ll 1, \label{L<1}
\end{equation}
then, to the lowest order of  $L(t)$,  one  finds  from
(\ref{V-szer}) \cite{9,10}
\begin{equation}
V_{\parallel}(t) \cong V_{\parallel}^{(1)} (t) \stackrel{\rm
def}{=} -i \int_{0}^{\infty} K(t - \tau ) \exp {[} i ( t - \tau )
PHP {]} d \tau . \label{V||=approx}
\end{equation}

Note that from the definition of $L(t)$ it follows that
\begin{equation}
L(t) \rightarrow 0 \;\;\;\;{\rm as} \;\;\;\; t\rightarrow 0, \label{L-0}
\end{equation}
which means that the condition (\ref{L<1}) is always fulfilled for $t \rightarrow 0$ and thus
$V_{||}(t)$ given by formula (\ref{V||=approx}) describes very well properties of
the subsystem under considerations for $t\rightarrow 0$.

\subsection{$n$--dimensional case}
Now let us consider a general case of $n$--dimensional subspace
${\cal H}_{||}$. Vectors from such subspaces describe states of
$n$--level ($n$--particle) subsystems. The only problem is to
calculate $P\exp [itPHP]$ in (\ref{V||=approx}) for the case of
$\dim ({\cal H}_{\parallel}) = n$. Note that it is convenient to
consider such ${\cal H}_{||}$  as the subspace spanned by a set of
orthonormal vectors  $\{ |{\bf e}_{j}\rangle{\}}_{j=1}^{n} \in
{\cal H}$, $\langle{\bf e}_{j}|{\bf e}_{k}\rangle ={\delta}_{jk}$.
Then the projection operator $P$ defining this subspace
can be expressed as follows
\begin{equation}
P = \sum_{j=1}^{n} |{\bf e}_{j}\rangle\langle{\bf e}_{j}|.
\label{P-n}
\end{equation}

The operator $PHP$ is selfadjoint, so the $(n \times n)$ matrix
representing $PHP$ in the subspace ${\cal H}_{\parallel}$ is
Hermitian matrix. Solving  the eigenvalue problem for this matrix,
\begin{equation}
PHP |{\lambda}_{j}\rangle = {\lambda}_{j} |{\lambda}_{j}\rangle,
\; \; ({\scriptstyle j =1,2,\ldots,n}), \label{l-j}
\end{equation}
one obtains the eigenvalues ${\lambda}_{j} =
{\lambda}_{j}^{\ast}$, and eigenvectors $|{\lambda}_{j} \rangle$,
$(j =1,2,\ldots,n)$. Here for simplicity we assume that
${\lambda}_{1} \neq {\lambda}_{2}  \neq \ldots \neq {\lambda}_{n}
\neq {\lambda}_{1}\neq \ldots $, etc.. In other words it is
assumed that all $\lambda_{j}$ are nondegenerate and  thus all $
|{\lambda}_{j} \rangle$ must be orthogonal,
\begin{equation}
\langle{\lambda}_{j} |{\lambda}_{k} \rangle = \langle{\lambda}_{j}
|{\lambda}_{j} \rangle  \, {\delta}_{jk}, \; \; ({\scriptstyle j,k
=1,2,\ldots, n}). \label{l-jk}
\end{equation}
By means of these eigenvectors one can define new projection
operators,
\begin{equation}
P_{j} \stackrel{\rm def}{=} \frac{1}{\langle{\lambda}_{j}
|{\lambda}_{j} \rangle}
|{\lambda}_{j}\rangle\langle{\lambda}_{j}|, \; \; ({\scriptstyle j
=1,2,\ldots, n}). \label{P-j}
\end{equation}
The property (\ref{l-jk}) of the solution of the eigenvalue
problem for $PHP$ considered implies that
\begin{equation}
P_{j} P_{k} = P_{j} {\delta}_{jk}, \; \; ({\scriptstyle j
=1,2,\ldots, n}), \label{P-jk}
\end{equation}
and that the completeness requirement for the subspace ${\cal
H}_{\parallel}$
\begin{equation}
\sum_{j=1}^{n} P_{j} = P, \label{copm}
\end{equation}
holds. Now, using the projectors $P_{j}$ one can write
\begin{equation}
PHP = \sum_{j=1}^{n} {\lambda}_{j} P_{j}, \label{PHP-sp}
\end{equation}
and
\begin{equation}
P e^{\textstyle +itPHP} = P \sum_{j=1}^{n} e^{+it {\lambda}_{j}
}P_{j}. \label{ex-PHP-sp}
\end{equation}

This last relation is the solution for the problem of finding
$P\exp [itPHP]$ in the considered case of nondegenerate
$\lambda_{j}$ and together with the formula (\ref{V||=approx})
for $V_{\parallel}(t)$ yields
\begin{equation}
V_{\parallel}^{(1)} (t) = - \sum_{j=1}^{n} PHQ
\frac{e^{\textstyle{-it(QHQ - \lambda_{j})}} - 1}{QHQ -
\lambda_{j}} QHP\,P_{j}, \label{V||(t)-l-n}
\end{equation}
which leads to $V_{||} \stackrel{\rm def}{=} \lim_{t \rightarrow
\infty} V_{||}^{(1)} (t)$,
\begin{equation}
V_{\parallel} = - \sum_{j=1}^{n} \Sigma ({\lambda}_{j}) P_{j},
\label{V-n-fin}
\end{equation}
where
\begin{equation}
\Sigma ( \epsilon ) = PHQ \frac{1}{QHQ - \epsilon - i 0} QHP.
\label{r24}
\end{equation}
This solves the problem of finding the effective Hamiltonian
\begin{equation}
H_{||} \equiv PHP + V_{||}, \label{H_||}
\end{equation}
(where $V_{||} = \lim_{t \rightarrow \infty} V_{||}(t)$) governing
the time evolution in the $n$--state subspace ${\cal
H}_{\parallel}$ of the total state space $\cal H$.

The simplest case is when the operator $PHP$ has $n$--fold
degenerate eigenvalue $\lambda_{0}$, that is when $\lambda_{1} =
\lambda_{2} = \dots = \lambda_{n} \stackrel{\rm def}{=}
\lambda_{0}$. Then
\begin{equation}
V_{\parallel}^{(1)} (t) = - PHQ \frac{e^{\textstyle{-it(QHQ -
\lambda_{0})}} - 1}{QHQ - \lambda_{0}} QHP, \label{V||(t)-l0}
\end{equation}
which gives
\begin{equation}
V_{||} = - \Sigma (\lambda_{0}). \label{V-l-0}
\end{equation}

The most interesting cases seem to be the cases when the
eigenvalues $\lambda_{j}$ of $PHP$ are $k$--fold degenerate, where
$k <n$. Then the form of $V_{||}$ differs from (\ref{V-n-fin}) and
(\ref{V-l-0}).

So, let $\lambda_{1}, \lambda_{2}, \ldots, \lambda_{k}$ be the
nondegenerate eigenvalues for $PHP$ and $\lambda_{k+1} =
\lambda_{k+2} = \dots = \lambda_{n} \stackrel{\rm def}{=}
\lambda$. Then
\begin{equation}
PHP = \sum_{j=1}^{k} {\lambda}_{j} P_{j} + \lambda (P -
\sum_{j=1}^{k} P_{j}) , \label{PHP-sp-l}
\end{equation}
(here $P_{j}$ is given by the formula (\ref{P-j})) and
\begin{equation}
P e^{\textstyle +itPHP} = P \sum_{j=1}^{k} e^{+it
{\lambda}_{j}}P_{j} + P (P - \sum_{j=1}^{k} P_{j})
e^{\textstyle{it\lambda}}. \label{ex-PHP-sp-l}
\end{equation}
Using this last relation and the general formula
(\ref{V||=approx}) for $V_{||}(t)$ and then taking $t \rightarrow
\infty$ one finds
\begin{equation}
V_{\parallel} = - \sum_{j=1}^{k} \Sigma ({\lambda}_{j}) P_{j} -
\Sigma (\lambda)(P - \sum_{j=1}^{k} P_{j}). \label{V-nk-fin}
\end{equation}

\subsection{$2$--dimensional case}

Let us pass on to  $n =2$ case, i.e. to the case of
two--dimensional subspace ${\cal H}_{||}$, which can be
applied to neutral meson complexes. So, if $ |{\bf e}_{j}\rangle = |{\bf j}\rangle, \;\;(j=1,2)$,
then the  projector $P$ is defined by
\begin{equation}
P \equiv |{\bf 1}\rangle\langle{\bf 1}| + |{\bf
2}\rangle\langle{\bf 2}|. \label{P}
\end{equation}
In the LOY approach it is assumed that vectors $|{\bf 1}\rangle$,
$|{\bf 2}\rangle$ considered above are eigenstates of
$H^{(0)}$ for a 2--fold degenerate eigenvalue $m_{0}$:
\begin{equation}
H^{(0)} |{\bf j} \rangle = m_{0} |{\bf j }\rangle, \; \;  j = 1,2,
\label{b1}
\end{equation}
where $H^{(0)}$ is the so called free Hamiltonian, $H^{(0)} \equiv
H_{strong} = H - H_{W}$, and $H_{W}$ denotes weak and other
interactions which are responsible for transitions between  the
eigenvectors of $H^{(0)}$, i.e., for the decay process.

If $H$ has the following property
\begin{equation}
PHP \equiv m_{0}\, P, \label{P-H12=0}
\end{equation}
that is for $ H_{12} = H_{21} = 0$,
the  approximate formula (\ref{V||=approx}) for $V_{\parallel}(t)$
leads to the following form of $P e^{itPHP}$,
\begin{equation}
P e^{\textstyle{ i t PHP}} = P e^{\textstyle{itm_{0}}},
\label{exp-PHP-H}
\end{equation}
and thus to
\begin{equation}
V_{\parallel}^{(1)} (t) = - PHQ \frac{e^{\textstyle{-it(QHQ -
m_{0})}} - 1}{QHQ - m_{0}} QHP, \label{V||(t)-H0}
\end{equation}
which leads to
\begin{equation}
V_{||} = \lim_{t \rightarrow \infty} V_{||}^{(1)} (t) = - \Sigma
(m_{0}). \label{V-H-0}
\end{equation}
This means that in the case (\ref{P-H12=0})
\begin{equation}
H_{||} = m_{0} \, P - \,\Sigma (m_{0}), \label{H||-H12=0}
\end{equation}
and  $H_{||} = H_{LOY}$.

On the other hand, in the case
\begin{equation}
H_{12} = H_{21}^{\ast} \neq 0, \label{H12n0}
\end{equation}
the form of $Pe^{itPHP}$ is much more complicated.

In the general case (\ref{H12n0}) one finds the
following expressions for the matrix elements  $v_{jk}(t
\rightarrow \infty ) \stackrel{\rm def}{=} v_{jk}$  of
$V_{\parallel}$ \cite{9,10},
\begin{eqnarray}
v_{j1} = & - & \frac{1}{2} \Big( 1 + \frac{H_{z}}{\kappa} \Big)
{\Sigma}_{j1} (H_{0} + \kappa ) - \frac{1}{2} \Big( 1 -
\frac{H_{z}}{\kappa} \Big)
{\Sigma}_{j1} (H_{0} - \kappa )\nonumber   \\
& - & \frac{H_{21}}{2 \kappa} {\Sigma}_{j2} (H_{0} + \kappa )
+ \frac{H_{21}}{2 \kappa} {\Sigma}_{j2} (H_{0} - \kappa ) ,
\nonumber \\
&  & \label{v-jk}\\
v_{j2} = & - & \frac{1}{2} \Big( 1 - \frac{H_{z}}{\kappa} \Big)
{\Sigma}_{j2} (H_{0} + \kappa ) - \frac{1}{2} \Big( 1 +
\frac{H_{z}}{\kappa} \Big)
{\Sigma}_{j2} (H_{0} - \kappa ) \nonumber  \\
& - & \frac{H_{12}}{2 \kappa} {\Sigma}_{j1} (H_{0} + \kappa ) +
\frac{H_{12}}{2 \kappa} {\Sigma}_{j1} (H_{0} - \kappa ) ,\nonumber
\end{eqnarray}
where $j,k = 1,2$, $H_{z} = \frac{1}{2} ( H_{11} - H_{22} ) $,
$H_{0} \stackrel{\rm def}{=} \frac{1}{2} ( H_{11} + H_{22} )$ and
$\kappa = ( |H_{12} |^{2} + H_{z}^{2} )^{1/2}$.
Hence, by (\ref{H||=def}), $h_{jk} = H_{jk} + v_{jk} $.
It should be emphasized that  all components  of the expressions
(\ref{v-jk}) are of the same order with respect  to $\Sigma (
\varepsilon )$.

Formulae (\ref{v-jk}) for matrix elements $v_{jk}$ become much simpler if  $ |H_{12}| \ll |H_{0}|$. Also the symmetries of the system lead to simpler form of $v_{jk}$ (see eg. system of neutral mesons in which CPT--symmetry is assumed to hold \cite{ijmpa1998}).

\subsection{1--dimensional case}

This is the simplest case. Here $|{\bf e}_{1}\rangle \equiv |{\alpha}\rangle $,
and $P_{1} = |\alpha\rangle\langle \alpha| $. This leads to
\begin{equation}
H_{||}(t) = h_{WW}(t)\, P_{1}, \;\;\; P_{1}HP_{1} = E_{1}\,P_{1},
\;\;\; V_{||}(t) = v_{WW}(t)\,P_{1},
\label{H-WW}
\end{equation}
where $E_{1} = \langle \alpha|H|\alpha \rangle$ and
\begin{equation}
v_{WW}(t) =
v_{Ww}^{(1)} (t) = - \langle \alpha |HQ \frac{e^{\textstyle{-it(QHQ -
E_{1})}} - 1}{QHQ - E_{1}} QH|\alpha \rangle. \label{V-WW(t)-E1}
\end{equation}
Thus
\begin{equation}
v_{WW} = \lim_{t \rightarrow \infty} v_{Ww}^{(1)} (t) = - \Sigma_{1}
(E_{1}), \label{v-WW}
\end{equation}
which gives the WW effective Hamiltonian $h_{ww}$ appearing in
(\ref{WW-1}).

\section{One--particle effective Hamiltonian at long time region}

From (\ref{WW-1}) one can conclude that the exact one--dimensional
effective Hamiltonian, $h(t)$, fulfils the following identity
(see \cite{pra,urbanowski-2-2009})
\begin{equation}
h(t) \equiv i \,\frac{1}{a(t)}\,\frac{\partial a(t)}{\partial t}, \label{h(t)}
\end{equation}
where
\begin{equation}
a(t) \equiv \langle \alpha |e^{\textstyle{-itH}}|\alpha \rangle  \equiv
\int_{Spec.(H)} \omega({ E})\;
e^{\textstyle{-i\,{ E}\,t}}\,d{ E}, \label{a(t)}
\end{equation}
is the survival amplitude.

In general, in the case of quasi--stationary  states it is
convenient to express $a(t)$ in the following form %$a(t) = a_{exp}(t) + a_{non}(t) $,
\begin{equation}
a(t) = a_{exp}(t) + a_{non}(t), \label{a-exp+}
\end{equation}
where $a_{exp}(t)$ is the exponential part of $a(t)$, that is
$a_{exp} = N\, \exp\,[-it(E_{\alpha}^{0} - \frac{i}{2}\,\gamma_{\alpha}^{0})]$,
($E_{WW}^{0}$ is the energy of the system in the unstable state $|\alpha\rangle$
measured at the canonical decay times (when the exponential decay law is valid),
$N$ is the normalization constant), and $a_{non}(t)$ is the non--exponential
part of $a(t)$. For times $t \sim \tau$, $|a_{exp}(t)| \gg |a_{non}(t)| $.

\begin{figure}[h]
\begin{center}
{\includegraphics[height=50mm,width=110mm]{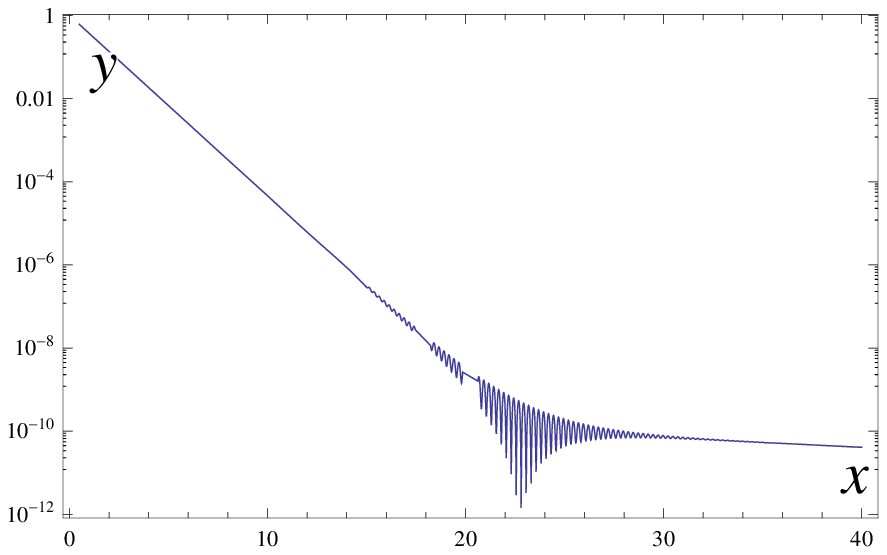}}\\ %50
\caption{Axes: $y={\cal P}(t)=|a(t)|^{2}$  --- the logarithmic scale,
$x = t / \tau$. ${\cal P}(t)$ is the survival probability.
The case  $\frac{E_{\alpha}^{0}}{\gamma_{\alpha}^{0}} = 25$.}
\label{p1}
\hfill\\
{\includegraphics[height=50mm,width=110mm]{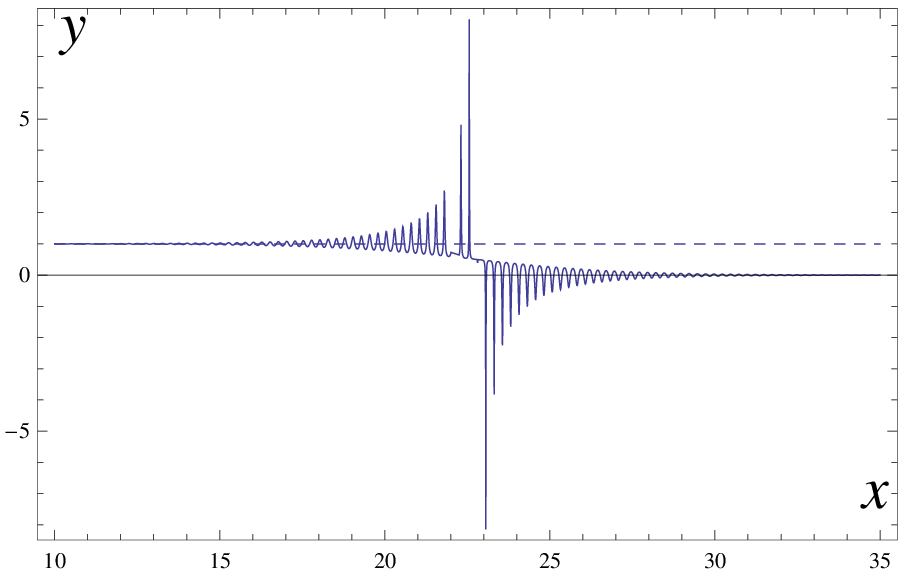}}\\
\caption{Fluctuations of $E_{\alpha}(t) =  \Re{\,[h(t)]}$
   at the transition time region.  Axes: $y= E_{\alpha}(t) / E_{\alpha}^{0}$,
   $x =t / \tau$.   $E_{\alpha}^{0} = \Re{\,[h_{WW}]}$,
   $\gamma_{\alpha}^{0} = -2\,\Im{\,[h_{WW}]}$. The case
   $\frac{E_{\alpha}^{0}}{\gamma_{\alpha}^{0}} = 25$.}
\label{F1}
\end{center}
\end{figure}

The transition time (or the crossover time) $t_{as}$
can be found by solving the following equation,
\begin{equation}
|a_{exp}(t)|^{\,2} = |a_{non}(t)|^{\,2}. \label{t-as}
\end{equation}
Long time properties of the survival probability ${\cal P}(t)=|a(t)|^{2}$
and the instantaneous energy ${ E}_{\alpha}(t) = \Re\,[h(t)]$ of the system
in the  unstable state $|\alpha\rangle$ are relatively
easy to find analytically for times $t \gg t_{as}$ even in
the general case (see \cite{urbanowski-1-2009}).
%as it was shown in \cite{urbanowski-1-2009}.
It is much
more difficult to analyze these properties  in the transition time region
where $t \sim t_{as}$. Typical forms of $P(t)$   and  $E_{\alpha}(t)$
are  presented in Figs (\ref{p1}) and  (\ref{F1}) respectively.
Results presented in these figures were obtained numerically by means of
the symbolic and numeric package {\em "Mathematica"}: Using  integral representation
(\ref{a(t)}) of $a(t)$ the amplitude $a(t)$ was found numerically for a given
$\omega(E)$, and then $|a(t)|^{2}$ and $\Re\,h(t)$ for $h(t)$ defined by (\ref{h(t)}).
Calculations were performed for
$\omega ({E}) = \frac{N}{2\pi}\,  \it\Theta ({ E}) \
\frac{\gamma_{\alpha}^{0}}{({ E}-{ E}_{\alpha}^{0})^{2} +
(\frac{\gamma_{\alpha}^{0}}{2})^{2}} $, $E_{min}=0$.

Methods used in the asymptotic analysis allow one to find a form of
$a(t) \equiv A_{\alpha \alpha}(t)$, (\ref{cmst-rho}), (\ref{a(t)}),
for large $t$ for all densities $\omega ({ E}) = \omega_{\alpha \alpha}(E)$,
corresponding with the case $(Spec.(H)\; = \;[E_{min}, +\infty)\,)$, for
which the Fourier transform (\ref{cmst-rho}), (\ref{a(t)}) exists.
For example, these calculations show that the amplitude $a_{non}(t)$ exhibits inverse
power--law behavior at the late time region: $t \gg t_{as}$.
The same can be done for the derivative of $a(t)$ and then using (\ref{h(t)})
a general asymptotic form of $h(t)$ can be found. It looks as follows
\begin{equation}
{h (t)\vline}_{\,t \rightarrow \infty} \simeq {E}_{min} + (-\,\frac{i}{t})\,c_{1} \,
+\,(-\,\frac{i}{t})^{2}\,c_{2} \,+\,\ldots, \label{h-infty-gen}
\end{equation}
where $ c_{i} = c_{i}^{\ast},\;\;i = 1,2,\ldots$
(compare \cite{urbanowski-1-2009}). This last result means that
\begin{equation}
 \Re{\,[h(t)]} \rightarrow E_{min}\;\;{\rm as}\;\;t\rightarrow \infty, \;\;\;{\rm and}\;\;\;
\Im{\,[h(t)]} \rightarrow 0 \;\;{\rm as}\;\;t\rightarrow \infty.
\end{equation}

\section{Final remarks}

The question arises: Can the effects described in Sec. 3, i.e., those presented in Figs (\ref{p1}) and  (\ref{F1})
and those following  from the relation (\ref{h-infty-gen}) be observed?
As it was mentioned earlier,
the effect presented in Fig (\ref{p1})
was confirmed experimentally \cite{rothe}.
This means that effects presented in Fig (\ref{F1}) and following
from (\ref{h-infty-gen}) have to take place too.

In general, there is a chance to observe some of
unstable particles, say $\phi$, which survived at $t \sim t_{as}$ only if there
is a source creating these particles in ${\cal N}_{\phi}^{\,0}$
number  such that
\begin{equation}
{\cal N}_{\phi}(t_{as}) =   {{\cal P}(t)\,\vline }_{\;t \sim
t_{as}}\;{\cal N}_{\phi}^{\,0} \; \gg \;1. \label{N(t-as)}
\end{equation}
From (\ref{t-as}) it follows that ${\cal P}(t_{as})
\simeq \exp\,[-\,\gamma^{0}_{\phi}\,t_{as}]$. This means that if there is a source creating
${\cal N}_{\phi}^{\,0} \gg \exp[\,+\,\gamma^{0}_{\phi}\,t_{as}]$ unstable particles
at the initial instant $t=t_{0} = 0$,
then a sufficiently large number ${\cal N}_{\phi}(t_{as}) $ of unstable particles $\phi$  has
to survive up to time $t_{as}$ or latter, and then the effect presented in Fig (\ref{F1})
should be observed. So in order to observe such effects
one needs an unstable  system having short $t_{as}$ (as it was used in the experiment
described in \cite{rothe}) or one should find sources creating sufficiently large number
${\cal N}_{\phi}^{\,0} $ unstable particles. Such sources are
known from cosmology and astrophysics and the these effects  should manifest itself there.

Effective Hamiltonians obtained in Sec. 2 are much more general and more
accurate than LOY and WW approximations. Formulae for $V_{\parallel}^{(1)}(t)$
and thus for $H_{\parallel}(t) = PHP + V_{\parallel}^{(1)}(t)$ derived in Sec. 2
seem to be a useful tool for a sufficiently accurate description of early
time properties of complexes of unstable particles evolving in time. They work
well for arbitrary $n < \infty$.

At times $t$ of order the lifetime $\tau$ the
time evolution of these complexes is well described  by $V_{\parallel}
\stackrel{\rm def}{=} \lim_{t \to \infty}\,V_{\parallel}^{(1)}(t)$ obtained
in Subsections 2.2 --- 2.4. Unfortunately these approximate formulae for
$H_{\parallel}(t)$ are unable to describe correctly very late time  properties
of complexes of unstable particles, when $t \sim t_{as}$ or $t >t_{as}$. It can
be done using the exact effective Hamiltonian of a form analogous to the
one--dimensional effective Hamiltonian $h(t)$ given by (\ref{h(t)}) and
using methods described in Sec. 3. Such a Hamiltonian acting in
$n$--dimensional subspace has the following form
\begin{equation}
H_{\parallel}(t) = i\,\frac{\partial \mathbb{A}(t)}{\partial t} \;
[ \mathbb{A}(t) ]^{-1}, \label{h(t)-n}
\end{equation}
where $\mathbb{A}(t) = \big[A_{\alpha \beta}(t)\big]$ is $(n\times n)$ matrix,
$\alpha,\,\beta\,= 1,2,\ldots,n$ and $A_{\alpha \beta}(t)$ are given by
(\ref{cmst-rho}). Asymptotically late time properties of matrix elements
$h_{\alpha \beta}(t)$ of this $H_{\parallel}(t)$ can be found using (\ref{cmst-rho}) and  applying
methods of asymptotic analysis to matrix elements $A_{\alpha \beta}(t)$
of $\mathbb{A}(t) $ and to $\frac{\partial}{\partial t}\,A_{\alpha \beta}(t)$.

\end{document}